\documentclass[conference]{IEEEtran}

\usepackage[T1]{fontenc}
\usepackage{algorithm}
\usepackage{algorithmic}
\usepackage{blindtext, graphicx}
\usepackage{balance}
\usepackage{amsmath}
\makeatletter
\newcommand*{\rom}[1]{\expandafter\@slowromancap\romannumeral #1@}
\makeatother

\begin{document}

\title{Efficient 3D Aerial Base Station Placement Considering Users Mobility by Reinforcement Learning}

\author{\IEEEauthorblockN{Rozhina Ghanavi\IEEEauthorrefmark{1},
    Elham Kalantari\IEEEauthorrefmark{2},
    Maryam Sabbaghian\IEEEauthorrefmark{1},
    Halim Yanikomeroglu\IEEEauthorrefmark{3}, and 
    Abbas Yongacoglu\IEEEauthorrefmark{2}}
  \IEEEauthorblockA{\IEEEauthorrefmark{1}School of Electrical and Computer Engineering\\
    University of Tehran, Tehran, Iran, Email: \{rghanavi, msabbaghian\}@ut.ac.ir}
  \IEEEauthorblockA{\IEEEauthorrefmark{2}School of Electrical Engineering and Computer Science\\
    University of Ottawa, Ottawa, ON, Canada, Email: \{ekala011, yongac\}@uottawa.ca}
  \IEEEauthorblockA{\IEEEauthorrefmark{3}Department of Systems and Computer Engineering\\
  Carleton University, Ottawa, ON, Canada, Email: halim@sce.carleton.ca}}

\maketitle

\begin{abstract}
This paper considers an aerial base station (aerial-BS) assisted terrestrial network where user mobility is taken into account. User movement changes the network dynamically which may result in performance loss. To avoid this loss, guarantee a minimum quality-of-service (QoS) and possibly increase the QoS, we add an aerial-BS to the network. For fair comparison between the conventional terrestrial network and the aerial-BS assisted one, we keep the total number of BSs identical in both networks. Obtaining the best performance in such networks highly depends on the optimal placement of the aerial-BS. To this end, an algorithm which can rely on general and realistic assumptions and can decide where to go based on the past experiences is required. The proposed approach for this goal is based on a discounted reward reinforcement learning which is known as Q-learning. Simulation results show this method provides an effective placement strategy which increases the QoS of wireless networks when it is needed and promises to find the optimum position of the aerial-BS in discrete environments.
\end{abstract}

\IEEEpeerreviewmaketitle

\section{Introduction}
Ubiquitous connectivity, reliable quality-of-service (QoS), extremely high number of connected devices, and supporting diverse  services are essentials of the fifth generation (5G) and beyond-5G (5G+) wireless networks. To obtain the distinctive attributes of 5G, it is imperative to exploit ingenious approaches. One of the emerging technologies which increases the agility of the network while enhancing the QoS, is utilizing aerial base stations (aerial-BSs), especially when the existing terrestrial infrastructure is insufficient to address the demand. Aerial-BSs can have several use cases, such as assisting the legacy terrestrial network to relieve congestion in case of an unusual excessive demand (such as a festival or a sport match). They can also provide temporal coverage when the ground-BSs are not accessible (such as a natural disaster which might damage the ground-BSs).
\let\thefootnote\relax\footnote{This work is supported in part by Huawei Canada Co., Ltd.}

\subsection{Related Work}
Use of aerial-BSs in wireless networks has gained attention in recent years. They may offer the most suitable way to enhance the QoS of the next generation wireless networks.
In \cite{h1, h2}, the positioning of aerial relays under assumption of fixed altitude is discussed.
In \cite{onnumber}, a novel approach for serving a number of users while using minimum number of aerial-BSs along with finding an efficient 3D placement is proposed.
In \cite{backhaulaware}, a backhaul aware 3D placement of an aerial-BS for various network design parameters is presented.
In \cite{efficientplacement}, an aerial-BS is positioned with the use of numerical methods to maximize the number of served user.
In \cite{fsobased}, a free space optical link is proposed as a method to provide backhaul/fronthaul access for flying network platforms.
In \cite{placementUAV}, the  optimal placement of an aerial-BS to maximize the number of users served with minimum transmit power is discussed.
In \cite{pir}, considering delay-tolerant and delay-sensitive users, an algorithm is proposed to find efficient 3D locations of aerial-BSs. The algorithm also investigates the user-BS associations and wireless backhaul bandwidth allocations to maximize the sum logarithmic rate of the users in a heterogeneous network including a macro-BS and several drone-BSs.
In \cite{DSC}, the downlink coverage performance of aerial-BSs is investigated.
In \cite{optimal}, the optimization problem of placing aerial-BSs and ground-BSs to minimize the average network delay is investigated.
In \cite{asso, distri}, aerial-BSs are considered as backhaul/fronthaul links and the association problem of small cells with aerial-BSs is investigated.
The effect of aerial relays between ground nodes in a wireless network is discussed in \cite{relay}.
 
\subsection{Our Contribution}
One of the drawbacks of the previous studies of aerial-BS systems is that none of them has considered users' movements. In other words, the optimization problem has been solved for a snapshot of the network. In reality, however, it is very likely that the position of the users changes gradually. Correspondingly, there is an indisputable link between the QoS of users and their movements. Another disadvantage of the previous studies is that the used optimization methods such as heuristic algorithms may need long processing time and this makes these methods impractical in real situations.

In this paper, we consider a traditional wireless network. User movement inevitably entails considerable changes in the QoS, raising the probability of the desired QoS not being delivered. We cope with this issue by exploiting an aerial-BS (Fig. 1). For a fair comparison between the aerial-BS assisted network and the terrestrial one, we keep the number of BSs in both networks identical.
As previously mentioned, the optimal placement of the aerial-BS can significantly influence the system performance. This issue has been the subject of several papers some of which are mentioned in Section \rom{1}.\textrm{A}. However, most of the articles considering this problem does not provide a closed form solution but present heuristic algorithms \cite{onnumber, pir}. In dynamic environments where the network topology changes, the heuristic algorithm needs to be reinitialized and run for the new topology. This process can be time consuming and imposes more computational complexity to the system. In our case where the network topology changes gradually due to users' movement, reinforcement learning is a promising candidate to solve the problem. Once the training phase is completed, this method continuously adapts the solution for small changes. This prevents the repeated running of the heuristic algorithm for continual topological changes. To the best of our knowledge, 3D positioning of aerial-BSs has not been solved using reinforcement learning. We propose an algorithm which significantly increase the QoS of the network and also it has lower complexity compare to other discussed method in this area.
\begin{figure}
  \includegraphics[width=0.5\textwidth]{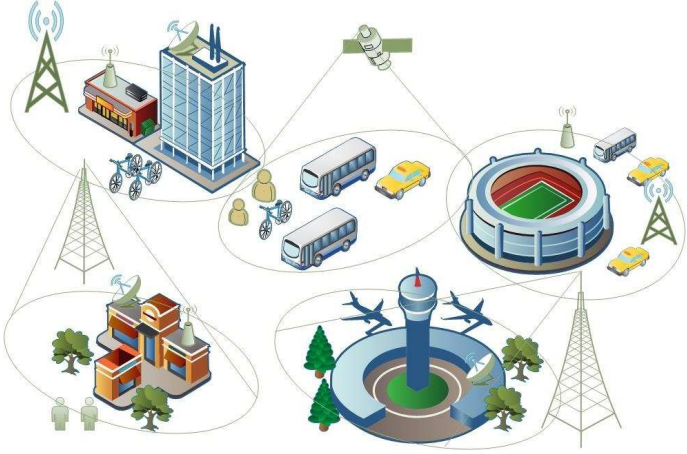}
  \centering
  \caption{Aerial-BS can assist network through hard to predict situations.}
\end{figure}

The rest of this paper is organized as follows. The system model is given in Section \rom{2}. Section \rom{3} gives the problem formulation and our novel approach to solve it. Section \rom{4} presents the simulation results and finally the study is concluded in Section \rom{5}.

\section{System Model}
We consider a downlink of a wireless cellular network including several macro-BSs and an aerial-BS. The parameters I and J denote the set of users and macro-BSs. We use $i=I\textsubscript{1},I\textsubscript{2},...,I\textsubscript{k},$ and $j=J\textsubscript{1},J\textsubscript{2},...,J\textsubscript{p},$ to index users and BSs, respectively. We assume that the users are served if the network QoS is higher than a threshold QoS which is denoted by QoS\textsubscript{th} throughout the paper. This parameter presents the QoS required by the network below which the quality is not acceptable. We assume that at the initial time of $t_{st}$, all users in the system fulfill this QoS requirement. However, as time passes, due to gradual movements of the users the QoS changes and it might fall below QoS\textsubscript{th}.
We propose a framework where by the use of an aerial-BS, we reduce the probability of losing the QoS of the users. Since the aerial-BS movement is costly, we cannot constantly change the position of the aerial-BS. In fact, our goal is to support at least the QoS\textsubscript{th} while the position of the aerial-BS is fixed at least for the minimum time interval of $t_{\textrm{min}}$. While users are moving in the area users' association to BSs change continuously. Our key solution is to include the prediction of users positions and calculate the users' association for each configuration in optimizing the location of the aerial-BS. This way, the QoS meets the required level without the cost being prohibitive.

\subsection{Air-to-ground Path Loss Model}
The air-to-ground path loss depends on the height of the aerial-BS and its elevation angle with the user. Several studies have investigated the air-to-ground path loss model. In this paper, we adopt the path loss model proposed in \cite{model1} and \cite{model2}. This model derives the air-to-ground path loss equation by considering two propagation classes, the first one is having a line-of-sight (LoS) links and the second one is the group of non line-of-sight (NLoS) links. The probability of having a LoS link is formulated as
\begin{equation}
P(\textnormal{LoS})=\frac{1}{1+\kappa\exp(-\zeta(\frac{180}{\pi})\theta-\kappa)},
\end{equation}
where the elevation angle of $\theta$ equals $\arctan(\frac{h}{l})$. The parameters $h$ and $l$ denote the altitude and the horizontal distance between the aerial-BS and the user, respectively. Constants $\kappa$ and $\zeta$ depend on the environment. The average path loss is defined as
\begin{equation}
\textsf{PL(dB)} = 20\log(\frac{4\pi f\textsubscript{c}d}{c})+ P(\textnormal{LoS})\eta\textsubscript{LoS}+P(\textnormal{NLoS})\eta\textsubscript{NLoS}.
\end{equation}
The first term of (2) presents the free space path loss which depends on the carrier frequency, $f_c$, the speed of light, $c$ and the distance between aerial-BS and user denoted by $d$. The constants $\eta\textsubscript{LoS}$ and $\eta\textsubscript{NLoS}$ present the additional loss due to free space propagation. We recall that $P(\textnormal{NLoS})$=1-$P(\textnormal{LoS})$.

\subsection{Users' Mobility Model}
As mentioned earlier, user movements can significantly affect the wireless network performance if the location of the aerial-BS is optimized only for a specific configuration. There are diverse methods to model the movement of ground users. The authors in \cite{modeluser} presented a comprehensive survey on such models. One of the most favored models is the random walk model. In this model the direction of the movement is determined by an angle uniformly distributed between $[0, 2\pi]$ and a user is assigned a random speed of a pedestrian between $[0,c_\textrm{max}]$, where $c_\textrm{max}$ denotes the maximum speed of a pedestrian. A graphical example of the random walk model is depicted in Fig. 2.
\begin{figure}
  \includegraphics[width=0.4\textwidth]{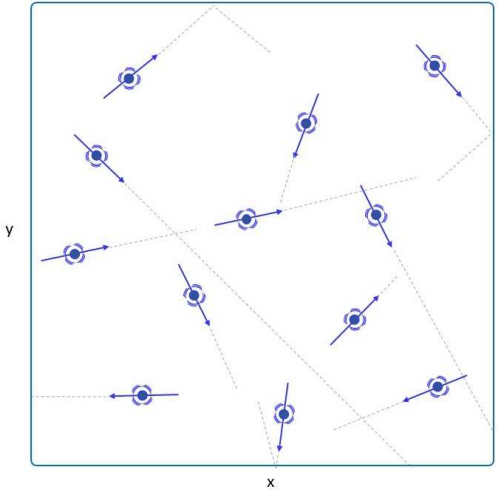}
  \centering
  \caption{Users' mobility scenario according to the random walk model.}
\end{figure}

\subsection{Optimization Problem}
To obtain the optimum position of the aerial-BS added to the system, we solve the following optimization problem.
\begin{equation}
\max_{h_{\chi},x_{\chi},y_{\chi}} 
\sum_{j=1}^{P}\sum_{i=1}^{k} {R_{ij}A_{ij}},
\end{equation}

subject to
\begin{equation}
\sum_{j=1}^{p} {A_{ij}}=1,
\end{equation}
\begin{equation}
h_\textrm{min} <h_{\chi}< h_\textrm{max},
\end{equation}
\begin{equation}
x_\textrm{min}<x_{\chi}< x_\textrm{max},
\end{equation}
\begin{equation}
y_\textrm{min} <y_{\chi}< y_\textrm{max},
\end{equation}
\begin{equation}
A_{ij} \in \{0,1\},
\end{equation}
where $A_{ij}$ is equal to 1 if the $i$-th user is assigned to the $j$-th BS and it is zero otherwise. The parameter $\chi$ is the assigned index to the aerial-BS, $[x_\textrm{min},x_\textrm{max}]$, $[y_\textrm{min},y_\textrm{max}]$, and $[h_\textrm{min},h_\textrm{max}]$ denote the 3D area which the aerial-BS could be placed in. In fact, we obtain the optimal position of the aerial-BS such that the aggregate network throughput is maximized. The term $R_{ij}$ denotes the throughput of the $i$-th user determined as
\begin{equation}
R_{ij}=\log_2 (1+\textnormal{SINR$\textsubscript{ij}$}),
\end{equation}
where signal to interference plus noise ratio, SINR, is defined as
\begin{equation}
\textnormal{SINR\textsubscript{ij}}=\dfrac{P\textsubscript{r{ij}}}{\sigma^2+\sum_{{k=1 \\ k\neq j}}^{p}I_{k}},
\end{equation}
where $P\textsubscript{r}$ is the received power and $\sigma^2$ is the noise power, the noise is a zero-mean white Gaussian noise with variance of $\sigma^2$. The term $I\textsubscript{k}$ illustrates the co-channel interference between the BSs. In this study, users choose the BSs based on the maximum SINR. In fact, as the users are moving along the area these assignments change and they decide which terrestrial or aerial-BS to connect based on the maximum SINR approach.

The problem of (3) is an NP-hard problem. Considering the fact that the environment is also changing and using heuristic approaches entails high computational complexity for repeated runnings, we solve this problem exploiting learning approaches. We utilize reinforcement learning specifically Q-learning as a novel approach to find the optimum position of the aerial-BS. It has been proved that under some constraints, this method can achieve the optimal solution in discrete environment if the learner has adequate time for learning \cite{proof}. Another advantage of this algorithm is that after sufficient learning time, the learning agent will learn the environment and will be able to find the optimum position in a very short time. This feature is essential to tackle the users' mobility problem, to the fact that this consideration makes the wireless network unsteady and it is necessary to have an algorithm which can guarantee a rapid solution to tune the wireless network with the new conditions.

\section{aerial-BS Positioning based on Q-learning}
Section \rom{3}.\textrm{A} presents an overview of reinforcement learning and Q-learning. Section \rom{3}.\textrm{B} applies this algorithm to the problem.
\subsection{Q-learning Algorithm}
Machine learning applications have gained attention in wireless networks in recent years. \cite{survey} has a survey on these works. Reinforcement learning has been widely applied in researches which need excessive sense of environments. It relies on the rewards and punishments received by a learning agent during the learning process. In the process of learning, the agent will learn taking the optimal procedure which in our case leads to maximizing the QoS. The learning agent, which is in the current state, takes an action and calculates the reward associated with taking that specific action. This continues until next action. Then, the state of the system changes and this process needs to be run on the new state. In each state transition, the agent generates a matrix including all information collected in state transitions. The information includes rewards and new states. This matrix is going to be used in later state transitions and helps the agent to improve the system performance further.

One of the most practical reinforcement learning techniques is Q-learning \cite{qlearning}, which does not need the exact transition formulation. This makes the method more realistic as the agent faces the current and actual wireless network not the formerly designed one.
There are number of studies related to the use of Q-learning in wireless networks. As an example, in \cite{proRL} the authors discuss this method to provide an efficient QoS for multimedia communication.
In a Q-learning algorithm, the agent considers a class of states $S=s\textsubscript{1},s\textsubscript{2},...,s\textsubscript{n},$ a class of actions $A=a\textsubscript{1},a\textsubscript{2},...,a\textsubscript{m},$ and a knowledge matrix $Q$. In each state, the learning agent performs an action, $a_{\iota}$ which triggers a state transition. Then, the agent calculates the reward in the new state. This procedure continues until the agent reaches to the desired goal and the algorithm converges \cite{suttonbook}.

\subsection{Proposed Algorithm}
In this paper, in a cellular system where users are moving, the goal is to maximize the QoS which is the aggregate network throughput (3). We attain this goal by adding an aerial-BS to the existing wireless network while users are moving with the random walk model.

We use Q-learning to find the optimal 3D position of the aerial-BS to increase the QoS in $t_{\textrm{min}}$. This goal is achieved by maximizing the reward $r_{\iota}$ in the state $s_{\iota}$ by taking the action $a_{\nu}$. In our learning process, we have events which are the movements of the users. If the QoS becomes lower than QoS\textsubscript{th}, aerial-BS needs to change its position. In this situation learning agent must decide which action it should take to change the state. We define 6 actions through which the aerial-BS can explore the plane. The area is between $[x_\textrm{min},x_\textrm{max}]$, $[y_\textrm{min},y_\textrm{max}]$, and $[h_\textrm{min},h_\textrm{max}]$ which are defined in Table \rom{2}. When the mentioned event occurs and the learner decides to take the desired action, $a_{\nu}$, the current position of the aerial-BS changes and the action takes the aerial-BS to a new position. Here, the states are defined as the positions that the aerial-BS can be in, in the 3D coordinate. The system will receive deterministic rewards based on the action the learner takes in each state. The reward in the $t$-th time interval is defined as
\begin{equation}
r_{t}=\textnormal{QoS$\textsubscript{t}$}-\textnormal{QoS$\textsubscript{t-1}$}.
\end{equation}
The $Q$ knowledge matrix for this problem consists of the following elements
\begin{equation}
Q(s_{t},a\textsubscript{t})=\alpha[r\textsubscript{t+1}+\gamma \max Q(s\textsubscript{t+1},a)-Q(s\textsubscript{t},a\textsubscript{t})],
\end{equation}
where $s_t$ is the state of the agent at time $t$ and $a_t$ is the corresponding action the agent takes.
$0<\alpha<1$ is the learning rate, which decrease throughout the learning process for the convergence of the solution. $0<\gamma<1$ is the discount factor which determines the speed of convergence and the accuracy of the process. In this study, we use $\epsilon-greedy$ strategy for choosing exploration or exploitation action in the learning process. Exploration in learning procedure allows the agent to neglect the locally optimal results. While, exploitation occurs when the agent takes optimum actions based on its knowledge.

At initial time $t=t_{st}$, we calculate the throughput of the system with existing ground-BSs and users and consider it as QoS\textsubscript{th}. Then users walk along the area with the random walk model. We assume that users continue walking in their decided direction for at least 10 s \cite{teraHz}.
In this process users' association to BSs change and the new QoS is calculated. While users are moving in the area Q-learning algorithm is running in order to find the optimum position for the aerial-BS. This calculation would be run while the system is being used. As wireless network usage continues, the time required in each iteration to obtain the optimal solution reduces. In each time when the system starts from $t_{st}$, we use the $Q$ matrix of the previous iteration as the initial $Q$ matrix. This increases the speed of convergence and can be considered as a tremendous motive to use Q-learning in the wireless network. Previous steps
explained earlier are presented in Algorithm 1.

\begin{algorithm}
\caption{Q-learning algorithm for 3D placement of an aerial-BS}
 \begin{algorithmic}[1]
 \renewcommand{\algorithmicrequire}{\textbf{Input:}}
 \renewcommand{\algorithmicensure}{\textbf{Output:}}
 \REQUIRE Distribution of BSs which are in hexagonal pattern and users with Poisson point process.
 \ENSURE  Optimum position of the aerial-BS
  \STATE QoS\textsubscript{th} is the calculated aggregate network throughput for the given ground wireless network.
  \WHILE {system is running}
  \STATE Let the users move for $t_{\textrm{min}}$ with the given users' mobility model.
  \STATE Calculate the aggregate network throughput for the new system.
  \IF {( QoS of the system with 19 ground-$\textrm{BSs} < \textnormal{QoS\textsubscript{th}}$)}
  \STATE Initialize $ Q(s,a) $ equal to the calculated $ Q(s,a) $ from the previous system running session.
  \FOR{ $\textrm{episode}= 1$ to max episode}
  \STATE Initialize $s$ as the previous state of the aerial-BS;
  \FOR { $\textrm{step=1}$ to max step}
  \STATE Decide which action to be taken, using $\epsilon-greedy$;
  \STATE Take action $a_{t}$, calculate the $r_{t}$, calculate the next $s$ of the aerial-BS;
  \STATE
   $Q(s\textsubscript{t},a\textsubscript{t})=\alpha[r\textsubscript{t+1}+\gamma \textnormal{ max}Q(s\textsubscript{t+1},a)-	Q(s\textsubscript{t},a\textsubscript{t})],$
    \STATE Update the state $s$ of the aerial-BS;

  \ENDFOR
  \ENDFOR
  \ELSE
  \STATE There is no need to use an aerial-BS.

  \ENDIF
  \ENDWHILE
 \end{algorithmic}
 \end{algorithm}
\section{Simulation Results}
We consider a cellular network with frequency reuse 1 in an urban area where 19 ground-BSs are placed in a hexagonal cell distribution. The number of users is between 150 and 800 and they are positioned based on Poisson point process. A random realization of users along with ground-BSs is presented in Fig. 3. The system parameters are presented in Table \rom{2}.

\begin{figure}[!t]
  \includegraphics[width=0.5\textwidth]{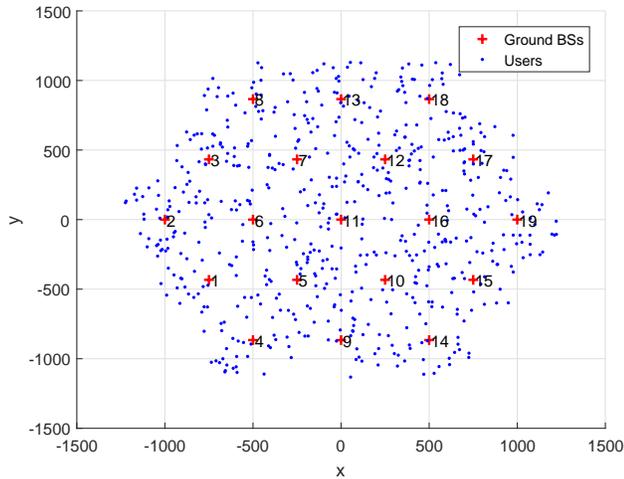}
  \centering
  \caption{2D distribution of users and ground-BSs at the beginning of the simulation ($t=t\textsubscript{st})$.}
\end{figure}

\begin{table}[t!]
  \renewcommand{\arraystretch}{1.3}
  \centering
  \caption{Urban Environment Parameters}\label{table1}
    \begin{tabular}[t]{|c |c |} 
      \hline
      \scriptsize\textbf{Parameter} & \scriptsize\textbf{Value} \\  
      \hline
      $\kappa$ & 9.61  \\ 
      \hline
      $\zeta$ & 0.16  \\
      \hline
      $\eta_{LoS}$ & 1 dB \\
      \hline
      $\eta_{NLoS}$ & 20 dB \\
      \hline
    \end{tabular}
  \bigskip
  \centering
  \caption{Simulation Parameters}\label{table2}
\begin{tabular}[t]{| c | c |} 
  \hline
  \scriptsize\textbf{Parameter} & \scriptsize\textbf{Value} \\ 
  \hline
$f_c$ & 2 GHz   \\
\hline
$c_\textrm{max}$ & 1.3 m/s  \\
\hline
$h\textsubscript{min}$ & 25 m\\
\hline
$h\textsubscript{max}$ & 525 m\\
\hline
Total 2D area & 4 km$^2$ \\
\hline
$t\textsubscript{min}$ & 10 s\\
\hline
$\gamma$ & 0.9\\
\hline 
$\epsilon$ & 0.9\\
\hline
\end{tabular}
\end{table}
Ground-BSs, new distribution of users and the 3D optimal position of the aerial-BS at t=100 s is shown in Fig. 4.
\begin{figure}
  \includegraphics[width=0.5\textwidth]{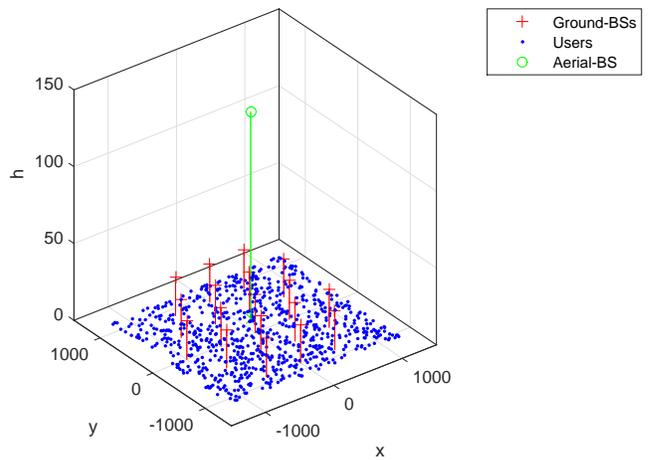}
  \centering
  \caption{3D distribution of users, ground-BSs and optimum position of aerial-BS at t=100 s.}
\end{figure}Fig. 5 presents the cumulative distribution function (CDF) for average SINR of users. This figure compares the CDF of average SINR for the traditional cellular system with 19 ground-BSs and our proposed system with 18 ground-BSs and supported by 1 aerial-BS. In the latter system, the position of the aerial-BS determined by Q-learning. This shows that overall, the given method improves the SINR parameter of system.
\begin{figure}[!t]
  \includegraphics[width=0.5\textwidth]{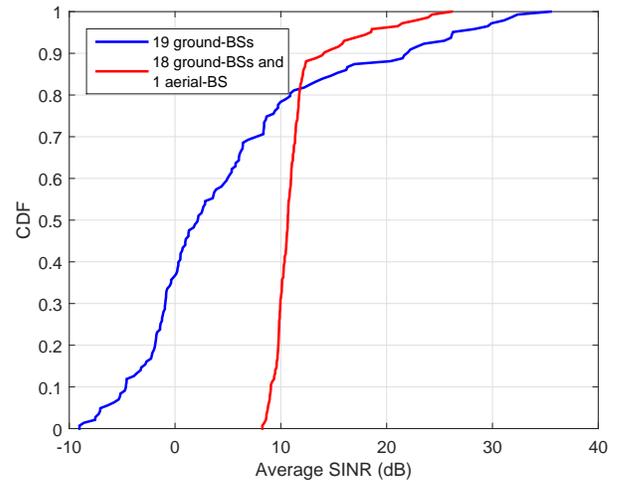}
  \centering
  \caption{CDF of average SINR of users for both the traditional ground-BSs system with 19 ground-BSs and our proposed system supported by 18 ground-BSs and 1 aerial-BS.}
\end{figure}The average spectral efficiency of both systems are presented in Fig. 6. As can be seen, the system assisted by aerial-BS outperforms the traditional ground-BS system in terms of the spectral efficiency.
\begin{figure}[!t]
  \includegraphics[width=0.5\textwidth]{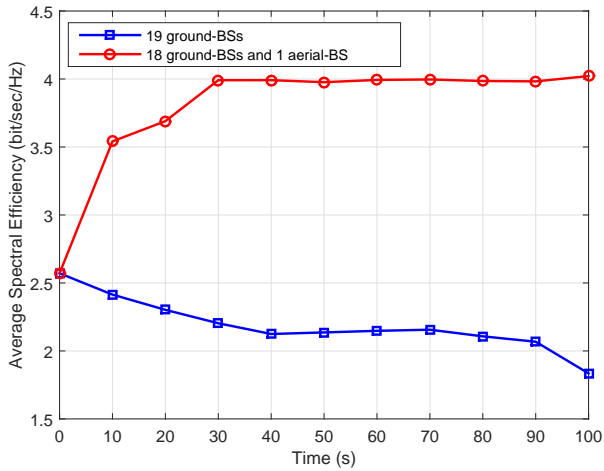}
  \centering
  \caption{The average spectral efficiency from both the traditional ground-BSs system with 19 ground-BSs and our proposed system supported by 18 ground-BSs and 1 aerial-BS.}
\end{figure}
Reward per iteration calculated from (11) for the presented wireless network with 18 ground-BSs and 1 aerial-BS is shown in Fig. 7. This figure shows the gradual improvement of the system as the reward reduces in each iteration, it also shows system approaches the optimum position after sufficient iterations and become stable as the reward becomes zero.
\begin{figure}[!t]
  \includegraphics[width=0.5\textwidth]{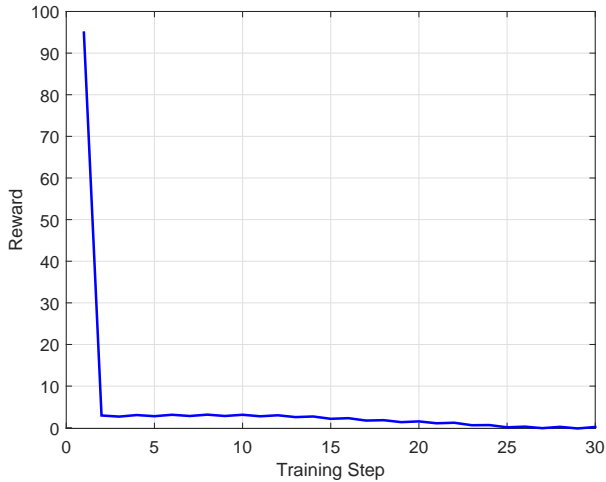}
  \centering
  \caption{Reward per iteration from solving (11), for finding the aerial-BS placement in a time slot.}
\end{figure}
\section{Conclusion}
In this paper, a traditional network of ground-BSs has been considered. To compensate the QoS loss due to user movements, the ground wireless network is assisted by an aerial-BS while one of the ground-BSs is switched off to keep the total number of BSs even and save energy. We apply Q-learning to the system as a novel approach to find the optimal position of the aerial-BS. Simulation results indicate that this method can bring much higher QoS to the network considering users' movements. By exploiting this approach, after giving the agent sufficient time to learn the environment, the processing time to find the optimum position of the aerial-BS becomes really low; therefore, it is a promising approach that can keep the agility and flexibility of future wireless networks. 

\balance

\end{document}